# THE EVALUATION CASE STUDY OF ONLINE COURSE DURING PANDEMIC PERIOD IN MONGOLIA

Uranchimeg Tudevdagva[1,2] , Bazarragchaa Sodnom[3] and Selenge Erdenechimeg[3]

[1]Faculty of Computer Science, Technical University of Chemnitz, Germany
[2]Power Engineering School, Mongolian University of Science and Technology, Mongolia
[3]Mongolian University of Pharmaceutical Sciences, Mongolia

### ABSTRACT

*This paper describes a test and case study of self-evaluation of online courses during the pandemic time. Due to the Covid-19, the whole world needs to sit on lockdown in different periods. Many things need to be done in all kinds of business including the education sector of countries. To sustain the education development teaching methods had to switch from traditional face-to-face teaching to online courses. The government made decisions in a short time and educational institutions had no time to prepare the materials for the online teaching. All courses of the Mongolian University of Pharmaceutical Sciences switched to online lessons. Challenges were raised before professors and tutors during online teaching. Our university did not have a specific learning management system for online teaching and e-learning. Therefore professors used different platforms for their online teaching such as Zoom, Microsoft teams for instance. Moreover, different social networking platforms played an active role in communication between students and professors. The situation is very difficult for professors and students. To measure the quality of online courses and to figure out the positive and weak points of online teaching we need an evaluation of e-learning. The focus of this paper is to share the evaluation process of e-learning based on a structure-oriented evaluation model.*

## 1. INTRODUCTION

The Mongolian University of Pharmaceutical Sciences (MUPS), the leading private university in pharmacy was founded in 2000 and it offers courses in Pharmacy, Traditional Medicine, Pharmaceutical Engineering and Technology, and Nursing. The Drug Research Institute and the GMP-Pharmaceutical Industry are the main components of the MUPS, and it has the great advantage of training to conduct a combination of Training-Research-Industry [1]. Due to the global pandemic, it was moved on public alert in Mongolia by Resolution No. 62 of February 12, 2020, and Resolution No. 178 of November 11, 2020, of the Government of Mongolia, and classroom learning at Universities and Colleges were temporarily closed and continue to move to a form of online learning [2]. E-learning has expanded since last year in all kind of education. Starting from the second semester of 2019-2020, the MUPS moved from classroom learning to online learning. As well as the first semester of 2019-2020, a total of 3001 video lessons, 489 enclosed files, and 41056 online test databases were created and 18,534 hours of electronic lectures, seminars, and workshops were organized under the basis of the cloud program [2]. From here, a total of 36 subjects of 93 credit hours, including general academic, professional foundation, and specialization, are taught online to first to fifth-grade students in pharmacy. As of the first semester of 2020, a total of 1277 students are studying at the MUPS, of which 1062 are studying in the Pharmacy Department. Our study involved 552 students, including 52 male and 500 female students.

The “Vision-2050” long-term development policy of Mongolia, the medium-term development plan of the education sector of Mongolia 2021-2030, and the Government action plan set the goal of “creating equal opportunities for everyone to receive quality education and strengthening the system of equal inclusion” [3]. In the future, it is necessary to reform the legal environment of the education sector to live and work in the age of digital technology, and to develop a humane and ethical citizen with active social participation and continuous development potential. In order to implement these policy objectives, there is an urgent need to reform the legal environment in the education sector. In addition, for the first time, the draft law regulates global development trends, the electric transition in education, online learning, distance learning, and open educational resources to provide citizens with equal and accessible access to education. This is the first time that electric learning, technology, and content have been legalized. Mongolia needs to have a well-structured, integrated research university that combines teaching and research in the present and future when the world needs to improve its competitiveness, efficiency, and rapid development of communication and information technology. Consequently, we need to compete as a world-class university. Accordingly, the teacher must have the knowledge and ability to organize non-classroom, mixed forms of training; to support the continuous development of the teaching profession at the national level by organizing pieces of training on teaching methods and professional skills development in a centralized, regional, distance and electronic form. In addition, communication, information technology infrastructure, software, and hardware for distance and electronic education are the responsibility of the relevant departments [4]. It has been a long time since the introduction of an online learning management system in the Mongolian education system. A common challenge for today's educational institutions is to manage online learning activities, disseminate training content, monitor student learning, motivate teachers and students in the learning process, and required the development of a learning environment to improve indicators, methodologies, and models for evaluating learning activities.

Although there are many examples and experiences of universities in developed countries using electronic learning systems, and it’s clear that the most important thing is to adapt these systems to the characteristics of the country's education system and to develop a new system that suits the country's education system and reflects the needs of educational institutions [5]. Due to the outbreak of Covid-19, a new coronavirus pandemic, classroom learning was banned from the spring semester of the 2019-2020 academic years. The introduction of fully online learning was a major challenge for the entire community, as many teachers who were not systematically trained in online learning techniques conducted the full course online for the first time throughout the semester. Each teacher went online and faced various problems, but each of them was able to solve the problem on their own and continue the training. However, in this article, we have selected our own assessment for a bachelor's degree in Pharmacy in order to evaluate our learning process, draw conclusions, and clarify our focus on future electric learning.

## 2. EVALUATION OF E-LEARNING

Evaluation of e-learning is one of the main points of discussion for researchers. Many researchers and educators developed different methods for the evaluation of e-learning. Successful e-learning can run as a result of collaboration where many different groups are working together: Professors,

Technical people, Administration, Tutors, Students, and others. Therefore evaluation of e-learning can be done in different complexity and level. As defined by Christopher Pappas, evaluation of e-learning was done before, during and end of e-learning. These are so-called assessment, formative and summative evaluations [6]. The e-learning and multimedia evaluations developed by Sage Learning Systems 17 major categories and 105 separate criteria on which e-learning courses can be applied for evaluation [7]. Prasad explains how an evaluation of e-learning should be done. By his idea understanding own value is important. Counting benefits of e-learning, convenient and flexible access, budget-friendly, measurable results, and reporting are key points of Prasad's view [8]. Petrone offers to use a famous method for evaluation from Kirkpatrick. He thinks that the Kirkpatrick model can be applied to measure the learning effectiveness of e-learning evaluation in the best way [9]. Donald Kirkpatrick developed an evaluation method for educational programs in four levels during his dissertation [10]. Level 1 focuses on reaction, level 2 focuses on learning, level 3 is for behavior and level 4 focuses on results. Kirkpatrick's four-level evaluation methodologies are well known and widely distributed model. Anstey and Watson established a method called "Rubric for E-Learning Tool" Evaluation. The main aim of this method is to identify the functionality and success of e-learning. They defined categories: functionality, accessibility, technical, mobile design, privacy, data protection and rights, social presence, teaching presence, cognitive presence. Each category consists of several covering spaces [11]. Bates described nine steps for the evaluation of online courses. Decide how you want to teach online is stands as the main question for the first step. Decide on what kind of online course stands for the second step. Work on a team defines the third step. Build on existing resources explains the fourth step. Master the technology describes the fifth step. Set appropriate learning goals are linked to the sixth step. Design course structure and learning activities are standing for the seventh step. Communicate, formulates the eighth step, and innovate explains the ninth step [12]. William Horton Consulting offers checklists for course comparison [13]. The quick and systematic way for evaluation is the key point of Horton. eLSE Methodology: a Systematic Approach to the e-Learning Systems Evaluation focuses on the most important aspects like Technology, Interaction, Content, Services [14]. The user-system interaction is an emphasizing aspect of the eLSE method. The newest approach in e-learning is to apply smart solutions based on artificial intelligence algorithms in medical teaching [15-16].

Evaluation of e-learning in Mongolia is not well developed till today. Before the pandemic period e-learning was more local issues of universities and evaluation of e-learning was not in the focus of accreditation institutions or the interest of professors. But some non-governance organizations (NGOs) put effort to develop national perspectives to the evaluation of e-learning early in 2012. Founder and directors of the NGO "E-Campus", Tsetseg-Ulzii cooperated with international quality assurance organizations together and published a report on criteria for evaluation of e-learning [17]. At the beginning of April of this year the UNESCO, United Nations Educational, Scientific and Cultural Organization reported on COVID-19 impact on education [18]. The report focused on the actual state of e-learning in Mongolia and its quality assurance issues. Some universities started to make decisions on university governance level focused on e-learning quality assurance and evaluation of online teaching. The Mongolian University of Science and Technology delivered new rules for e-learning management [19].

## 3. THE SURE MODEL

For self-evaluation of online courses, we selected the structure-oriented evaluation, SURE model. The SURE model was developed by the first author of the paper for the evaluation of e-learning. The SURE model consists of eight steps to evaluate the selected object [20-23].

Table 1. Eight steps of the SURE model

| Steps | Focus | Process |
|---|---|---|
| 1 | Definition of key goals | In this step evaluators should define key goals of evaluation. The SURE result will be positive, only if all defined key goals reach its target successfully. |
| 2 | Definition of sub goals | After definition of key goals evaluators should define sub goals which support in reaching key goals target. If any of sub goal reached its target then the corresponding key goal is evaluated as successful. |
| 3 | Confirmation of goal structures | Pre-defined key and sub goals structure should be discussed by all interested groups of evaluation results. Only after confirmation of goal structures the next step can be taken. If evaluators cannot accept the goal structures the first and second steps should be repeated until goal structures are accepted. |
| 4 | Preparation of questionnaire | Based on confirmed goal structures questions for the questionnaire should be prepared. |
| 5 | Confirmation of questionnaire | Prepared questionnaire should be checked and confirmed by all interested groups in evaluation results. Until full acceptance of all questions previous step should be repeated. |
| 6 | Collection of data | When questionnaire is ready data should be collected by objective ways. Best method to collect objective data is to use online survey. |
| 7 | Processing of data | The collected data should be processed by mathematical rules of the SURE model. There are exists an online calculator for the SURE evaluation [24]. |
| 8 | Reporting of results | The processed data will be shown via table from online calculator. Four main evaluation scores are available from table: General evaluation score, Evaluation scores for key goals, Evaluation scores for sub goals and Evaluation scores of each participants. |

To implement the SURE model we applied all eight steps in corresponding order.

## 3.1. The Evaluation of Online Course

For self-evaluation we selected Pharmacy courses of Bachelor students from MUPS. In course 1062 enrolled 552 students including 52 male and 500 female students.

## 3.2. Step From One to Three

These steps are focused on definition and confirmation of evaluation goals.

- Definition of key goals. Based on discussion of evaluators we defined four key goals (Figure 1):
    - Lecture quality ($B_1$);
    - Seminars/Practices/Laboratories quality ($B_2$);
    - Teaching performance ($B_3$);
    - Teacher skills for e-learning ($B_4$).

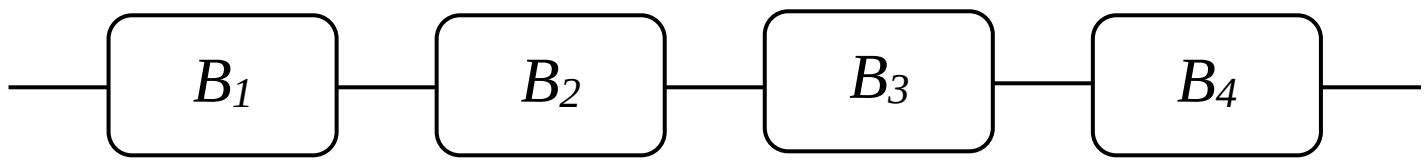

Figure 1. Key goal structure

- Definition of sub goals. To reach defined key goals here described sub goals (Figure 2):

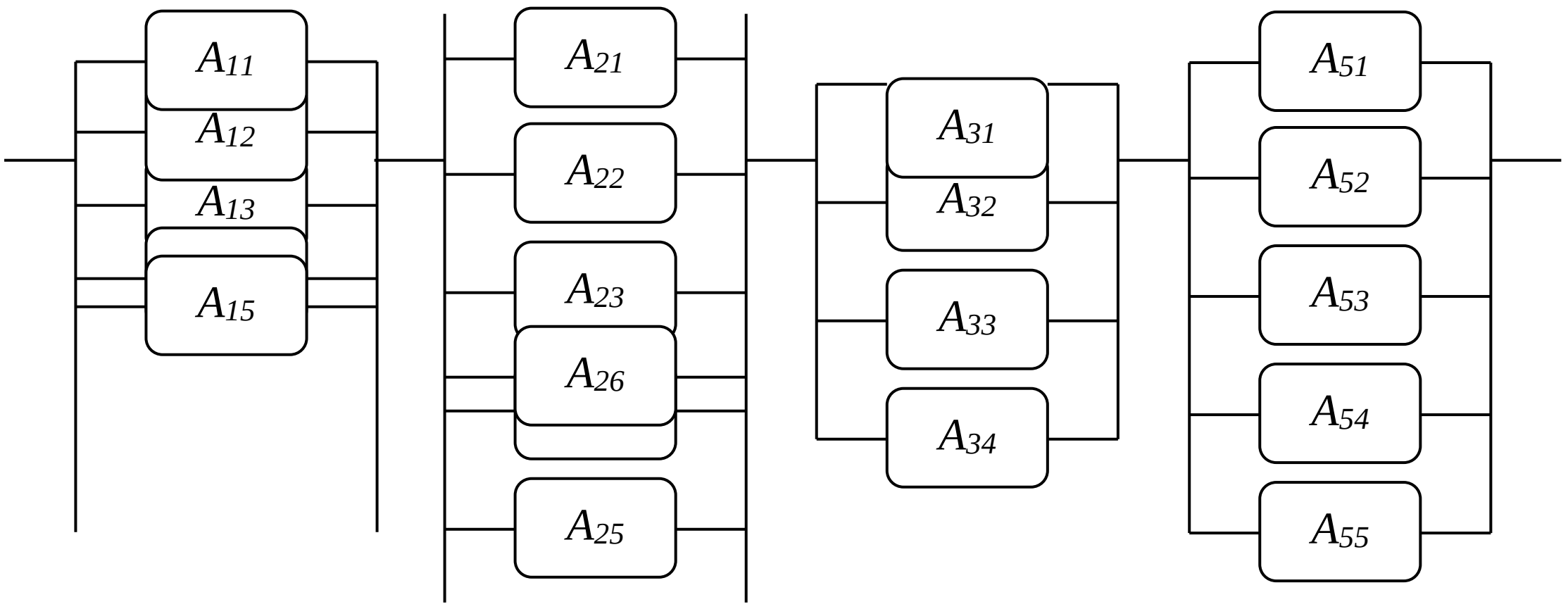

Figure 2. Sub goal structure

- Lecture quality ($B_1$):
  - Content of lecture ($A_{11}$);
  - Lecture lesson ($A_{12}$);
  - Time management of teacher during lecture ($A_{13}$);
  - Online synchronic lecture ($A_{14}$);
  - Offline video lecture ($A_{15}$).
- Seminars/Practices/Laboratories quality ($B_2$):
  - Knowledge ($A_{21}$);
  - Practical skills ($A_{22}$);
  - Learner centred learning ($A_{23}$);
  - Knowledge ($A_{24}$);
  - Online synchronic practice ($A_{25}$);
  - Offline video practice ($A_{26}$).
- Teaching performance ($B_3$):
  - Preparation to lesson ($A_{31}$);
  - Teaching methods ($A_{32}$;
  - Oral skills ($A_{33}$);
  - Technical skills ($A_{34}$).
- Teacher skills for e-learning ($B_4$):
  - Preparation level ($A_{41}$);
  - Files for lessons ($A_{42}$);
  - Motivation for students ($A_{43}$);
  - Communication with students ($A_{44}$);
  - Feedback speed and style ($A_{45}$).

- Confirmation of goals structures. The SURE evaluation expert and professors who teach courses confirmed the goal structures.

### 3.3. Step four and five

The focuses of these steps are preparation of the questionnaire. In step four questions based on sub goals structure have to be formulated. Scale for measurement designed from 0 to 4. 0 = Disagree at all, 1 = Up to 30% agree, 2 = 31-50% agree, 3 = 51-70% agree and 4 = 79-100% agree (Table 2).

Table 2. Example of question

| **№** | **Evaluation indicators** | **Disagree at all** | **Up to 30% agree** | **31-50% agree** | **51-75% agree** | **76-100% agree** |
|---|---|---|---|---|---|---|
| 1. | The content of the lecture was clear | | | | | |
| 2. | Lecture is comprehensible | | | | | |
| 3. | The teacher was able to use the class time perfectly for the lecture content | | | | | |
| 4. | It was convenient to teach the lecture online directly in accordance with the schedule | | | | | |
| 5. | It was convenient to watch the lecture on video and study it independently | | | | | |

Formulated questions should be checked and controlled by all interested groups of evaluation results. Only tested and accepted questionnaires can be used for data collection.

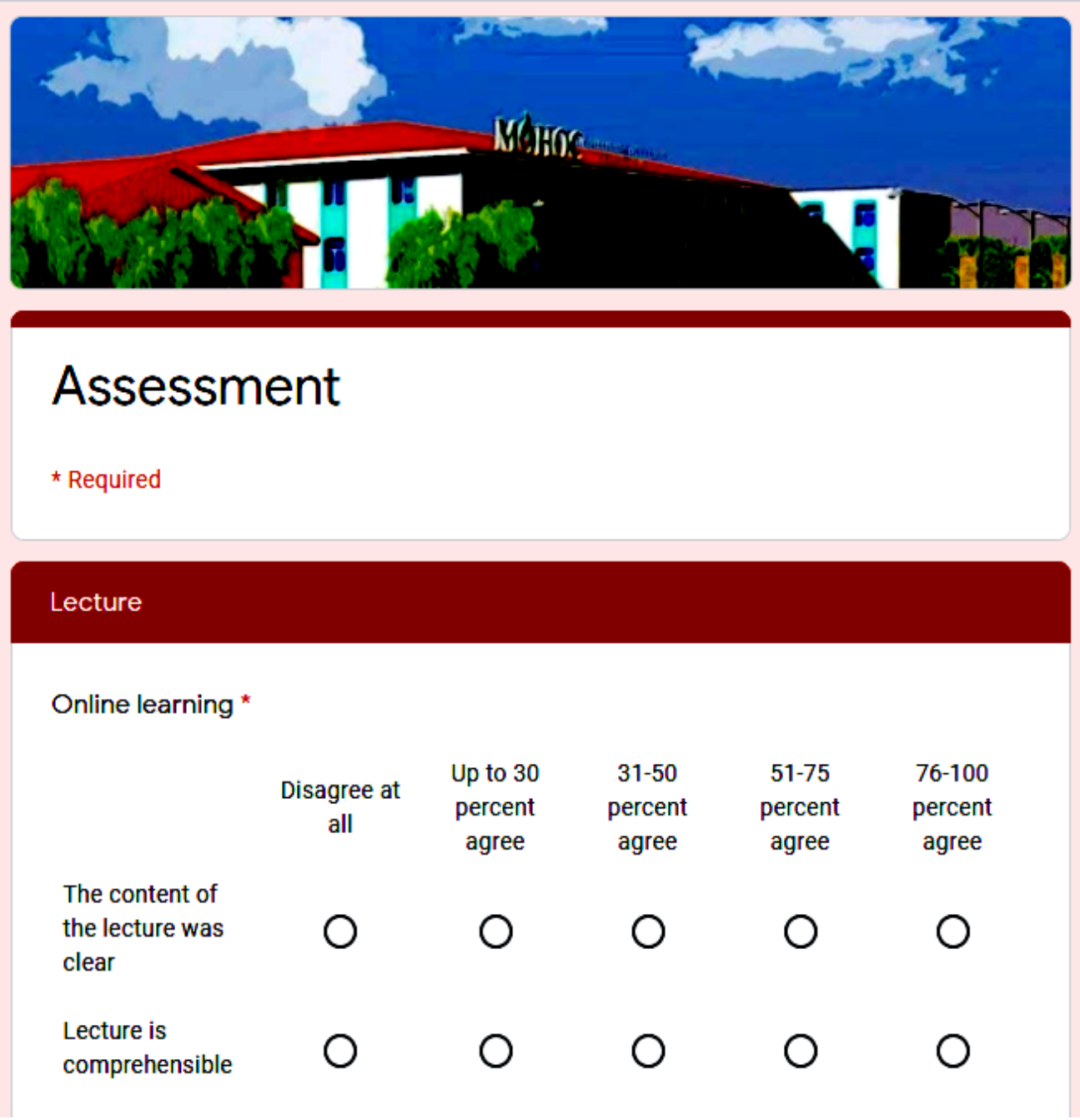

Figure 3. Online data collection

## 3.4. Step six

Data collection is a important part of any evaluation process. Objectively in good time collected data can increase probability of final evaluation score. For objective data collection we applied Google Form as online tool (Figure 3).

## 3.5. Step seven

Data which collected by Google form in original shows in Figure 4. The collected data should be processed by the SURE evaluation rules. Main mathematic rules defined based on goal structures [20]. To process the collected data (Figure 5) by online calculator the original data should be transferred to comma separated vector (CSV) files (Figure 6).

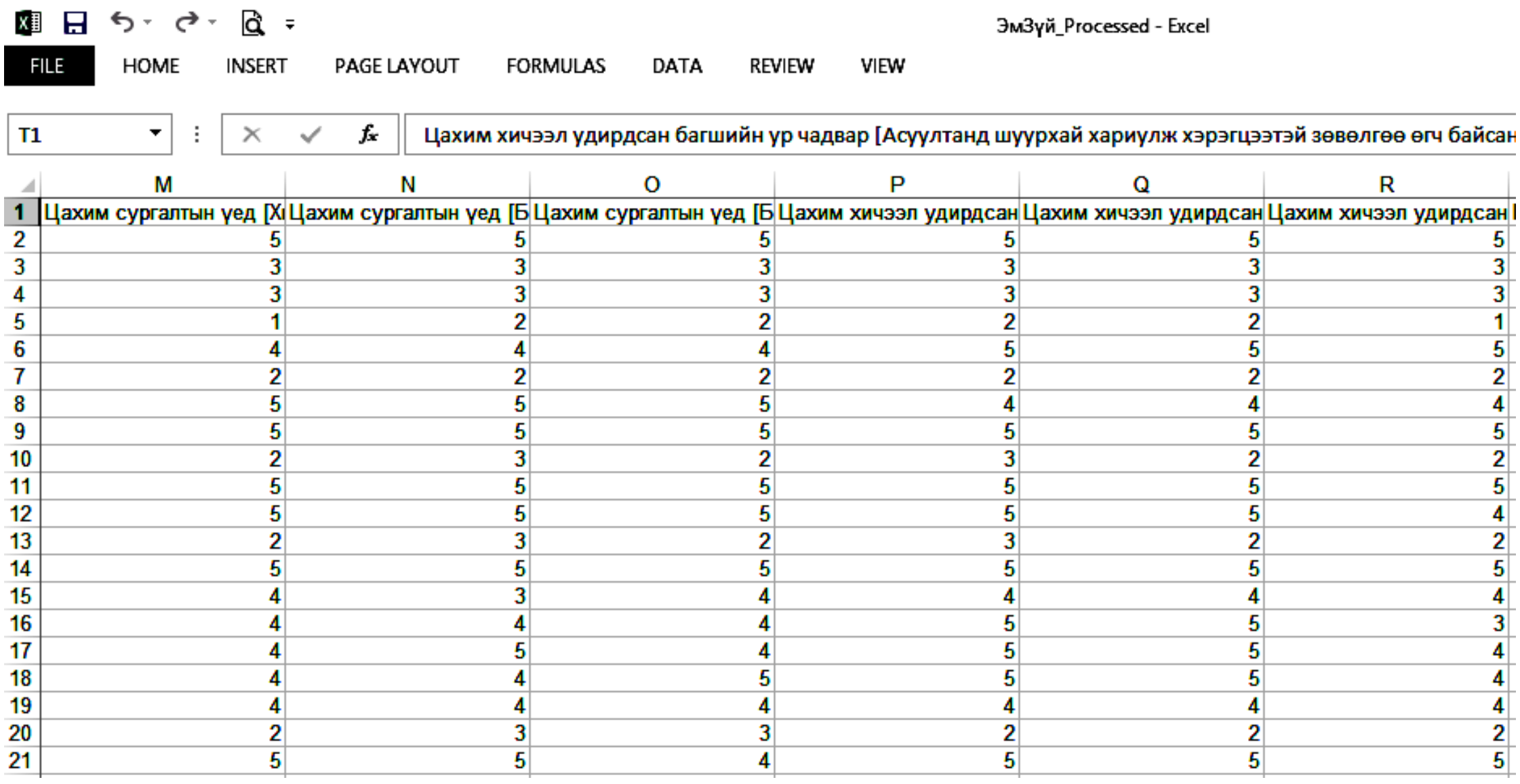

| | M | N | O | P | Q | R |
|---|---|---|---|---|---|---|
| 1 | Цахим сургалтын үед [Х | Цахим сургалтын үед [Б | Цахим сургалтын үед [Б | Цахим хичээл удирдсан | Цахим хичээл удирдсан | Цахим хичээл удирдсан |
| 2 | 5 | 5 | 5 | 5 | 5 | 5 |
| 3 | 3 | 3 | 3 | 3 | 3 | 3 |
| 4 | 3 | 3 | 3 | 3 | 3 | 3 |
| 5 | 1 | 2 | 2 | 2 | 2 | 1 |
| 6 | 4 | 4 | 4 | 5 | 5 | 5 |
| 7 | 2 | 2 | 2 | 2 | 2 | 2 |
| 8 | 5 | 5 | 5 | 4 | 4 | 4 |
| 9 | 5 | 5 | 5 | 5 | 5 | 5 |
| 10 | 2 | 3 | 2 | 3 | 2 | 2 |
| 11 | 5 | 5 | 5 | 5 | 5 | 5 |
| 12 | 5 | 5 | 5 | 5 | 5 | 4 |
| 13 | 2 | 3 | 2 | 3 | 2 | 2 |
| 14 | 5 | 5 | 5 | 5 | 5 | 5 |
| 15 | 4 | 3 | 4 | 4 | 4 | 4 |
| 16 | 4 | 4 | 4 | 5 | 5 | 3 |
| 17 | 4 | 5 | 4 | 5 | 5 | 4 |
| 18 | 4 | 4 | 5 | 5 | 5 | 4 |
| 19 | 4 | 4 | 4 | 4 | 4 | 4 |
| 20 | 2 | 3 | 3 | 2 | 2 | 2 |
| 21 | 5 | 5 | 4 | 5 | 5 | 5 |

Figure 4. Original data Excel sheet

...
5,5,5,5,5,5,5,5,5,5,5,5,5,5,5,5,5,5,5,5,
3,3,3,3,3,3,3,3,3,3,3,3,3,3,3,3,3,3,3,3,
3,3,3,2,3,3,3,3,3,3,3,3,3,3,3,3,3,3,3,3,
2,2,2,3,3,2,2,2,2,2,2,1,1,2,2,2,2,1,2,2,
4,4,4,4,4,4,4,4,4,4,4,4,4,4,4,5,5,5,5,5,
2,2,2,2,2,2,2,2,2,2,2,2,2,2,2,2,2,2,2,2,
5,5,4,5,5,5,4,5,4,5,4,5,5,5,5,4,4,4,5,5,
5,5,5,5,4,4,4,5,5,5,5,5,5,5,5,5,5,5,5,5,
3,2,3,2,1,3,2,2,3,2,1,2,2,3,2,3,2,2,1,2,
5,5,5,5,5,5,5,5,5,5,5,5,5,5,5,5,5,5,5,5,
...

Figure 5. Part of CSV file

The CSV file should be entered to online calculator of the SURE model (Figure 6).

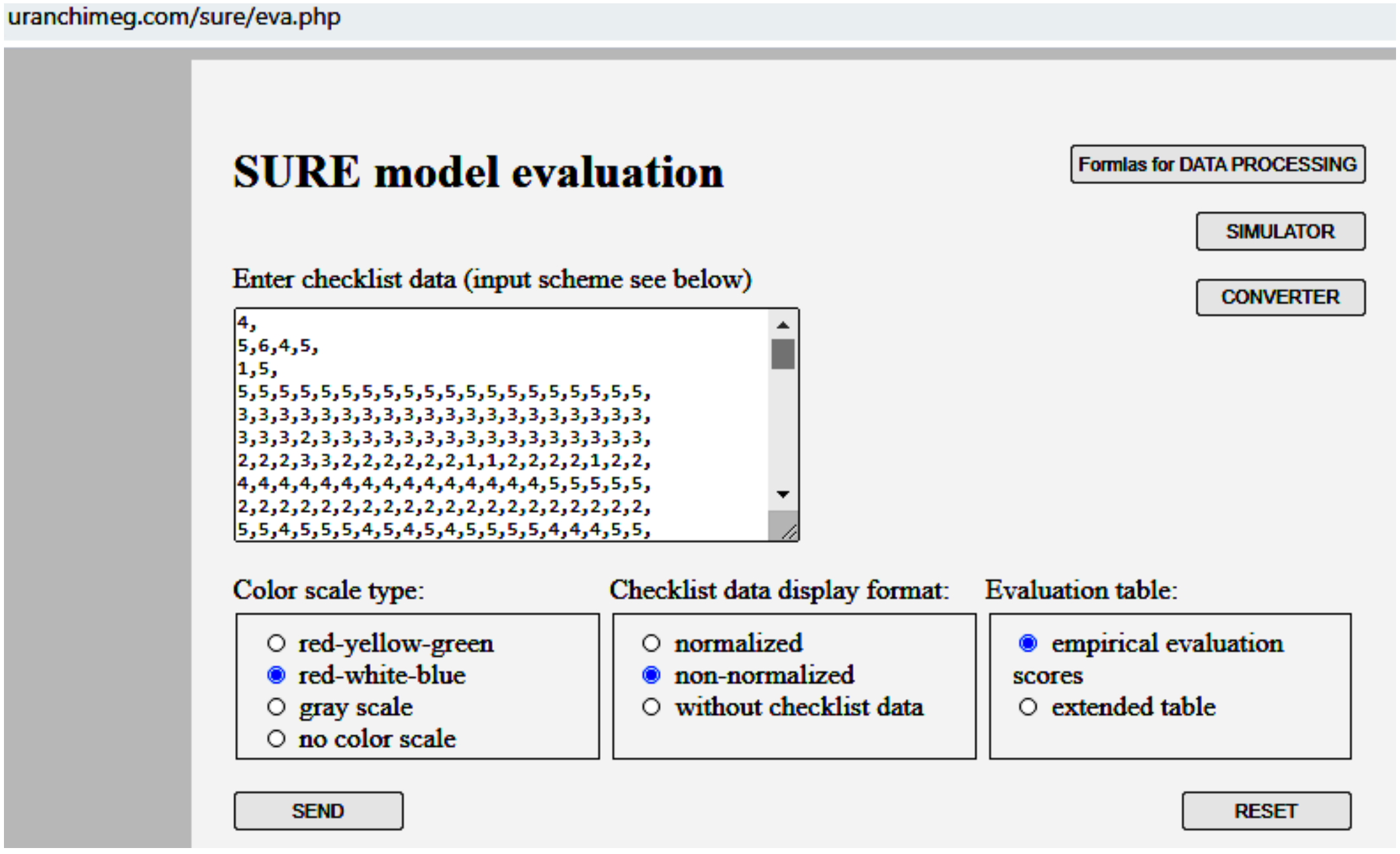

Figure 6. CSV file as enter data for online calculator

The given data will be processed by online tool and as outcome from online tool table with evaluation scores will be returned (Figure 7).

| | $B_1$ | | | | | $B_2$ | | | | | | $B_3$ | | | | $B_4$ | | | | | |
|---|---|---|---|---|---|---|---|---|---|---|---|---|---|---|---|---|---|---|---|---|---|
| | $A_{11}$ | $A_{12}$ | $A_{13}$ | $A_{14}$ | $A_{15}$ | $A_{21}$ | $A_{22}$ | $A_{23}$ | $A_{24}$ | $A_{25}$ | $A_{26}$ | $A_{31}$ | $A_{32}$ | $A_{33}$ | $A_{34}$ | $A_{41}$ | $A_{42}$ | $A_{43}$ | $A_{44}$ | $A_{45}$ | |
| $Q^*(A_{ij})$ | 0.69 | 0.66 | 0.68 | 0.63 | 0.65 | 0.64 | 0.61 | 0.62 | 0.62 | 0.61 | 0.59 | 0.65 | 0.68 | 0.72 | 0.71 | 0.73 | 0.73 | 0.66 | 0.67 | 0.69 | $Q_e^*(C) = 0.7022$ |
| $Q_e^*(B_i)$ | 0.75 | | | | | 0.68 | | | | | | 0.73 | | | | 0.75 | | | | | |

Figure 7. The evaluation result

## 3.6. Step Eight

The SURE model calculated four different evaluation scores.

- Main evaluation score. $Q^*_e(C)=0.70$
- Evaluation scores of key goals.
  - $B_1=0.75$;
  - $B_2=0.68$;
  - $B_3=0.73$;
  - $B_4=0.75$.
- Evaluation scores of sub goals.
  - $A_{11}=0.69$; $A_{12}=0.66$; $A_{13}=0.68$; $A_{14}=0.63$; $A_{15}=0.65$;
  - $A_{21}=0.64$; $A_{22}=0.61$; $A_{23}=0.62$; $A_{24}=0.62$; $A_{25}=0.61$; $A_{26}=0.59$;
  - $A_{31}=0.65$; $A_{32}=0.68$; $A_{33}=0.72$; $A_{34}=0.71$;
  - $A_{41}=0.73$; $A_{42}=0.73$; $A_{43}=0.66$; $A_{44}=0.67$; $A_{45}=0.69$;
- Evaluation scores of each response. We received 552 responses from students. Due to this big amount of data, the table with students' response data could not be included into the paper. Therefore, we noted here some results as example. 155 students evaluated the course with maximum score 1. 17 students evaluated with worst score 0.

### 3.7. Result of Evaluation

1062 students studied professional “Pharmacy” but only 552 students sent responses and it was 51.98% of total enrolled students. From them 500 were female and 52 were male students. Out of 97 students studying in a physician of traditional medicine, 60 students or 61.85% (50 female, 10 male), 25 students or 60.97% (17 female, 8 male) out of 41 students studying in Pharmaceutical Engineering and Technology, 56 students or 72.72% out of 77 students studying in Nursing % (54 female, 2 male) students took part in the evaluation process. The reason for some students not participating in the process was the hard accessibility to the Internet in rural areas.

The general evaluation score is 0.70 which is not so high, which is expected of professors who teach selected courses. Key goals are similar to general score: B1=0.75, B2=0.68, B3=0.73 and B4=0.75. There is no key goal that could reach over 80%. That means professors need to do a detailed analysis on each key goal and to focus on weak points of online teaching. The worst score of 0.59 received is “Offline video practice (A26)”. That means students cannot accept or adapt to studying from video by themselves. Further reasons can be not well-developed self-study ability of students or quality of video lessons could not meet expectations of students. Here we need a more detailed study to find out the reasons why this score is the worst and how it can be improved in the next courses. The best evaluation score 0.73 received are “Preparation level (A41)” and “Files for lessons (A42)” from the fourth key goal “Teacher skills for e-learning (B4)”. Students are satisfied with teaching skills for online teaching. Students respect the efforts of teachers for the preparation of online courses. However, 0.73 is not 1 and there are a lot of spaces for improvement of teacher skills for e-learning.

One of the most interesting results of the evaluation is 157 students evaluated online courses with a maximum score of 1. Against this 17 students evaluated online courses with the worst score of 0. Here, a big variance can be observed. This result shows more analyses on other questions relating to the learning environment, quality of internet connection speed, and other issues is needed to be done which can influence the evaluation to such an extent.

## 4. Discussions

Educational evaluation, especially evaluation of e-learning is a huge space for research and study. Therefore we need to discuss and reflect on existing evaluation models and methods, evaluation cases. Evaluators need to openly share their study and experiences on e-learning evaluation. There are several evaluation models and methods distributed among educational intuitions and accreditation organizations. But not all models and methods are suitable for all types of courses and e-learning. Most existing models and methods are fit to use at the university level, not for single online courses. The structure-oriented evaluation model tried to cover all nuances of the evaluation process. Beginning with the definition of evaluation goals until reporting of evaluation results included eight steps or the structure-oriented evaluation model. The second and third authors first time applied the SURE model for their case and it helped them to understand e-learning more deeply from the perspective of students.

There are can come to arise questions: need to do include professors or teaching staff into the evaluation process? Do we need self-evaluation for teaching courses? What is the aim of such activities? Which kind of evaluation methods or models should apply for self-evaluation? Do we need a standard question for self-evaluation of e-learning? and so on. It will be great if evaluators can unite their effort for such us study and cooperate more on same topics. Our opinion is self-evaluation is essentially important to understand the whole evaluation process and get a deeper feeling of online teaching. We will continue to do self-evaluations and share our experiences in the e-learning field.

## 5. CONCLUSIONS

The main aim of this paper is to share the evaluation case of online courses during the pandemic period at the Mongolian University of Pharmaceutical Sciences. This case study was the first test of the self-evaluation process for online courses by professors. Such kind of self-evaluation helps professors get a deeper understanding of the e-learning process, different roles, and functions of involved groups, expectations of students during online teaching. The cooperation with the SURE expert helped professors to understand the selected model and to do a successful self-evaluation. Professors collected experiences from the self-evaluation of e-learning by structure-oriented evaluation model. To learn to do a scientific self-evaluation for their online courses give opportunities to professors to look at detailed elements of e-learning from the inner structure of online teaching. Evaluation results show successful pars of online teaching and figure out weaknesses of e-learning. Such results are support professors to create a development plan for further teaching online.

We conclude this evaluation case study as the beginning point of a long-term study in direction of e-learning evaluation. In the future, the evaluation goals structure will be updated and the formulation of questions based on sub-goals will be improved. Moreover, we will distribute our case study to our colleagues not only from our university, but we will also share these experiences as well as with faculties from other universities, too. The most key conclusion is educators need to train with self-evaluations to learn more about their teaching methodologies during e-learning and to understand students' expectations from professors.

## AUTHORS

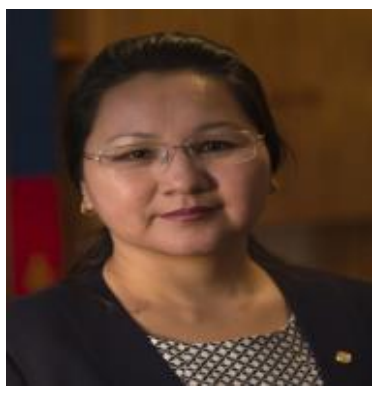

**Prof. Dr. Dr. h. c. Uranchimeg Tudevdagva**, Guest Professor of Faculty for Computer Science at Chemnitz University of Technology. Since 2010 until today holds the position of Professor of Power Engineering School at Mongolian University of Science and Technology. Prof. Tudevdagva is an expert on evaluation model for complex systems and e-learning.

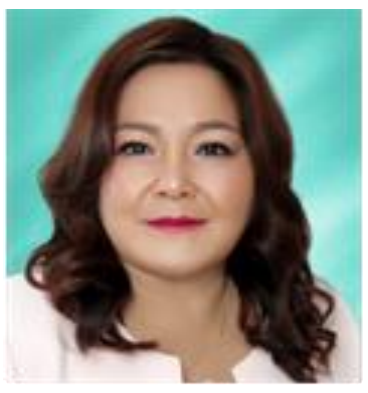

**Bazarragchaa Sodnom**

Dr. in Medical Science
Professor, Faculty of Traditional Medicine, Mongolian University of Pharmaceutical Sciences

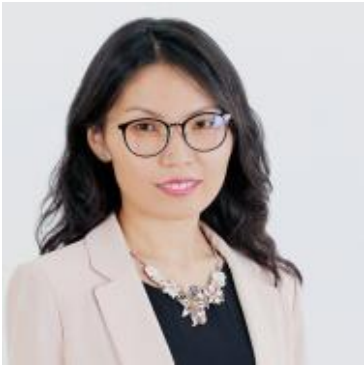

**Selenge Erdenechimeg**

Dr. in Pharmaceutical Sciences
Director of Graduate School
Mongolian University of Pharmaceutical Sciences, Mongolia
Research topic is: Phytochemical study of Mongolian medicinal plants